\documentclass[twocolumn,english,aps,prd,reprint,floatfix,notitlepage,footinbib,preprintnumbers,longbibliography]{revtex4-1}
\pdfoutput=1
\usepackage{lmodern}

\usepackage[T1]{fontenc}
\usepackage[latin9]{inputenc}
\usepackage{geometry}
\usepackage{subfigure,lmodern, amsmath,amssymb, graphicx, pifont, adjustbox, bm, xcolor}
\usepackage{amsfonts}
\usepackage{enumitem}
\usepackage{comment}
\usepackage{mathtools}
\usepackage{float}
\usepackage{slashed}
\usepackage{ragged2e}
\usepackage{array}
\usepackage{nameref}

\DeclareRobustCommand{\looongrightarrow}{%
  \DOTSB\relbar\joinrel\relbar\joinrel\relbar\joinrel\rightarrow
}

\usepackage{hhline}

\makeatletter\g@addto@macro\bfseries{\boldmath}\makeatother

\makeatletter\newcommand{\labeltext}[2]{%
  \def\@currentlabel{#1}%
  \label{#2}%
}
\makeatother

\usepackage{stackengine}
\usepackage{esint}
\usepackage[unicode=true,pdfusetitle,
 bookmarks=true,bookmarksnumbered=false,bookmarksopen=false,
 breaklinks=false,pdfborder={0 0 1},backref=false,colorlinks=true]
 {hyperref}
\hypersetup{
 pdfauthor={Grant N. Remmen},
 citecolor=black,linkcolor=black,urlcolor=black}

\newcommand{\appendixref}[1]{\hyperref[#1]{appendix~\ref{#1}}}
\def\equationautorefname~#1\null{eq.\,(#1)\null}
\usepackage{breakurl}
\usepackage[hang,flushmargin]{footmisc} 
\allowdisplaybreaks
\makeatletter

\usepackage{etoolbox}
\apptocmd{\thebibliography}{\justifying\setlength{\leftskip}{7.4mm}}{}{} 
 
 \usepackage{relsize}
\usepackage{babel}

\makeatletter
\def\simgt{\mathrel{\lower2.5pt\vbox{\lineskip=0pt\baselineskip=0pt
           \hbox{$>$}\hbox{$\sim$}}}}
\def\simlt{\mathrel{\lower2.5pt\vbox{\lineskip=0pt\baselineskip=0pt
           \hbox{$<$}\hbox{$\sim$}}}}
\makeatother

\usepackage{changepage}

\newcommand{\be}{\begin{equation}}
\newcommand{\ee}{\end{equation}}
\newcommand{\bea}{\begin{eqnarray}}
\newcommand{\eea}{\end{eqnarray}}

\newcommand{\mysec}[1]{\noindent {\bf #1.}---}
\newcommand{\mysubsec}[1]{\noindent {\it #1.}---}

\newcolumntype{P}[1]{>{\centering\arraybackslash}p{#1}}

\usepackage{fix-cm}

\begin{document}

\title{Zeta Functions and the Superstring}

\author{Grant N.~Remmen}
\affiliation{\scalebox{1}{Center for Cosmology and Particle Physics, Department of Physics, New York University, New York, NY 10003, USA}}    
    
\begin{abstract}

\noindent The superstring amplitude's Mellin transform in energy is computed at fixed momentum transfer.
In the forward limit, it is shown that this transformed amplitude reduces to the Riemann zeta function, while for $t\neq 0$ it represents a deformation of $\zeta$.
This object exhibits remarkable mathematical properties as a result of physical attributes of the string amplitude.
For integer subtractions, the dispersion relation defined by the Mellin transform yields the effective field theory expansion of the string amplitude order by order in $s$ at finite $t$, for which a new closed-form expression is derived.

\noindent 
\end{abstract}
\maketitle 

\mysec{Introduction}An enduring theme in theoretical physics is the discovery of unexpected connections and the recurrent appearance of mathematical objects in surprising places.
One such object is the Riemann zeta function $\zeta(z)$.
Celebrated for its role in the theory of the primes and defined via a Dirichlet series,
\be
\zeta(z) = \sum_{n=1}^\infty \frac{1}{n^z},
\ee
for ${\rm Re}(z)\,{>}\,1$, the zeta function appears in a vast array of contexts across mathematics and physics.
A century-old question is whether the Riemann hypothesis---which posits that all nontrivial zeros of $\zeta(z)$ take the form $z=\frac{1}{2}+i\tau$ for real $\tau$---has a physical justification~\cite{Montgomery,Polya,BerryKeating,Srednicki:2011zz,Bender:2016wob,Gutzwiller,Bhaduri:1994de,LeClair:2018cmo,LeClair:2023keb,Remmen:2021zmc}, as suggested by the Hilbert-P\'olya conjecture that these $\tau$ are the eigenvalues of some Hermitian operator.

A related question is whether the Riemann zeta function could show up in the S-matrix of some well-defined theory. The construction of scattering amplitudes with an infinite number of poles, by design closely connected to the Riemann zeta function, was explored in Ref.~\cite{Remmen:2021zmc}, with various mappings between the physical properties of these conjectured amplitudes and attributes of the zeta function.

However, it is even more interesting if we can locate $\zeta$ within a physical theory that we already know, so that its properties can be understood independently of a purpose-built amplitude construction.
A model with an S-matrix particularly suited to such exploration---containing various special functions, exhibiting an infinite tower of states, and with particularly nice behavior at high energies---is string theory.
While string amplitudes can be written in terms of ratios of zeta functions in a manner that does not depend on the nontrivial zeros~\cite{Freund:1987ck,He:2015jla}, and sequences of multiple zeta values appear in the low-energy effective field theory (EFT) expansion of the string~\cite{ZagierZerbini,Schlotterer:2012ny,Green:2019tpt}, in this paper we take the connection further, relating the full analytic structure of the Riemann zeta function to string scattering.

\begin{figure}[t]
\includegraphics[width=5.5cm]{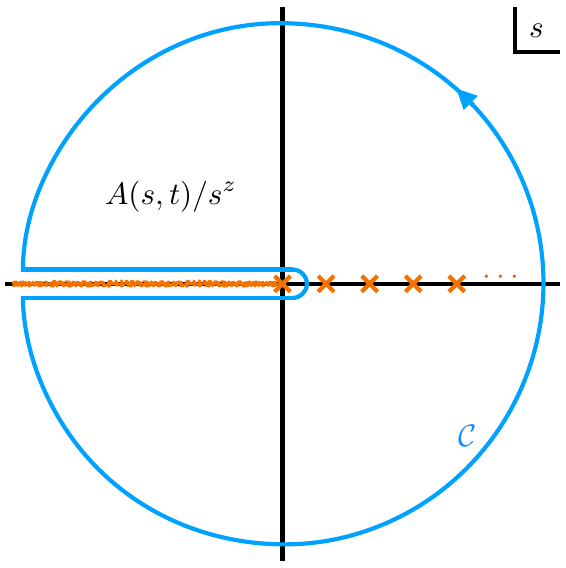}
\caption{Illustration of the contour integral defining $\Omega(z,t)$ in Eq.~\eqref{eq:contour}. The contour ${\cal C}$ in the $s$ plane comprises a circle at $s=\infty$, except for a Hankel contour that infinitesimally avoids the negative real $s$ axis, where the function $A(s,t)/s^z$ has a branch cut when $z$ is not an integer. The poles in $A(s,t)$ occur at nonnegative integer $s$.}
\label{fig:contour}
\end{figure}

Concretely, we will consider the four-gluon scattering amplitude of the open superstring~\footnote{Specifically, we are considering the polarization-stripped amplitude of the open superstring~\cite{Green:1981xx,Schwarz:1982jn}, which is similar to the Veneziano amplitude of the bosonic string~\cite{Veneziano:1968yb}, with its characteristic Euler beta function form. Of course, the Koba-Nielsen integral form of the string amplitude~\cite{Koba:1969rw,Arkani-Hamed:2024nzc} is a Mellin transform on the worldsheet variables, but here we are interested in Mellin transforming the Mandelstam variables of the amplitude itself.},
\be 
A(s,t) = -\frac{\Gamma(-s)\Gamma(-t)}{\Gamma(1-s-t)}, \label{eq:A}
\ee
where the Mandelstam variables $s$ and $t$ are, respectively, the squared energy of the initial two-particle state in the center-of-mass frame and the square of the exchanged momentum.
We compute the Mellin transform of the amplitude in $s$ at fixed $t$,
\be 
\Omega(z,t) = \frac{1}{2\pi i}\oint_{\cal C} \frac{{\rm d}s}{s^z} A(s,t),\label{eq:contour}
\ee
where ${\cal C}$ is the positively oriented keyhole contour in Fig.~\ref{fig:contour}, comprising a circle at infinity with a deformation around the negative real axis.
The $s^z$ factor produces a branch cut in the integrand on the negative $s$ axis for noninteger $z$, which the contour avoids.
While Mellin transforms of string amplitudes at fixed {\it angle} have been considered in the context of celestial amplitudes~\cite{Arkani-Hamed:2020gyp}, $\Omega(z,t)$ is instead defined at fixed $t$.

We will find that this $\Omega$ function has many remarkable mathematical properties, which are each in direct correspondence with the physical properties of the string amplitude. In particular, $\Omega(z,t)$ reduces to the Riemann zeta function in the forward limit of $t\,{\rightarrow}\,0$, and for general $t$ can be viewed as an interesting deformation of $\zeta(z)$.
We summarize the correspondence between the physical properties of the string amplitude and features of $\Omega$ that will be explored in this work as follows:

\begin{widetext}
\begin{center}
\begingroup
\renewcommand{\arraystretch}{1}
\begin{tabular}{c c c | c}
Properties of superstring amplitude $A(s,t)$ & $\big\vert$ & Properties of $\Omega(z,t)$ & $\;$Eq.$\;$ \\
\hline
Meromorphicity & $\looongrightarrow$ & Dirichlet series & \eqref{eq:sum}\\
Ultrasoftness & $\looongrightarrow$ & Zeros at $z=-k$ & \eqref{eq:integerzeros} \\
Unitarity & $\looongrightarrow$ & $\partial_t^k\Omega(z,t)>0$ for $z>0$ & \eqref{eq:unitarity} \\
Universality of $t=0$ string residue & $\;\;\looongrightarrow\;\;$ &  $\;\;\;$ Riemann zeta function at $t=0$ $\;\;\;$ & \eqref{eq:Omegazeta} \\
Residue duality $R(n,t)=R(t,n)$ & $\looongrightarrow$ & Finite sum of zeta functions at $t=k$ & \eqref{eq:fullposiden} \\
Level truncation & $\looongrightarrow$ & $\;$ Finite sum of exponentials at $t=-k$ $\;$ & \eqref{eq:leveltruncation} \\
Level truncation + residue duality & $\looongrightarrow$ & Shift relation & \eqref{eq:shift}
\end{tabular}
\endgroup
\end{center}
\end{widetext}

\medskip
\mysec{Properties of $\Omega$}Since the string amplitude is meromorphic, with poles at nonnegative integer $s$, we can deform the contour to write $\Omega$ as a Dirichlet series,
\be
\Omega(z,t) = \sum_{n=1}^\infty \frac{R(n,t)}{n^z}.\label{eq:sum}
\ee
The residues on the poles of $A(s,t)$ at $s=n$ are
\be
R(n,t) = \lim_{s\rightarrow n}(n-s)A(s,t)=\frac{\Gamma(n+t)}{\Gamma(n+1)\Gamma(t+1)}. \label{eq:R}
\ee
Since $A(s,t)\sim s^{t-1}$ in the Regge limit of large $s$ and fixed $t$, for negative $t$ we have highly convergent ultraviolet behavior known variously as superpolynomial softness~\cite{Haring} or ultrasoftness~\cite{Cheung:2025tbr}: for any power law $\propto s^{-k}$, there is a range of $t$ for which $A(s,t) \lesssim s^{-k}$ at large $s$. As a result, the sum in Eq.~\eqref{eq:sum} converges for ${\rm Re}(z)>{\rm Re}(t)$. Of course, just as in the case of the Riemann zeta function, we will be able to analytically continue $\Omega(z,t)$ to the entire $z$ plane at fixed $t$, modulo poles.

Remarkably, the enhanced ultraviolet behavior of string amplitudes implies the existence of zeros of $\Omega(z,t)$ as follows. If $z = -k$ for some integer $k>0$, the branch cut in Fig.~\ref{fig:contour} does not contribute, and the simple pole in $A(s,t)$ at $s=0$ is removed by the $1/s^z$ integrand, allowing us to deform the contour defining $\Omega(z,t)$ so that it equals just the pole at infinity in Eq.~\eqref{eq:contour}. The $A\sim s^{t-1}$ Regge scaling of the string amplitude implies that for $k < -{\rm Re}(t)$ this pole at infinity vanishes. An infinite tower of sum rules is therefore satisfied by the string residues for $t$ in this range,
\be
\sum_{n=1}^{\infty} n^k R(n,t) = 0,\label{eq:sumrules}
\ee
which one can verify using Eq.~\eqref{eq:R}. We thus have an infinite sequence of zeros of $\Omega(z,t)$ at $z = -k > {\rm Re}(t)$,
\be
\Omega(-k,t) = 0,\label{eq:integerzeros}
\ee
as a consequence of the string amplitude sum rules following from ultrasoftness.

Meanwhile, quantum mechanical unitarity implies that the residues of the string amplitude have a positive partial wave expansion.
As a consequence, one famously has that $\lim_{t\rightarrow 0}\partial_t^k R(n,t) \geq 0$ for all $k\geq 0$ due to the positivity of derivatives of Legendre polynomials in the forward limit.
Hence, unitarity implies positivity for $\Omega(z,t)$ and its derivatives,
\be
\lim_{t\rightarrow 0} \partial_t^k \Omega(z,t) > 0\;\;\;{\rm for}\;\;\;z>0.\label{eq:unitarity}
\ee

\medskip

\mysubsec{Identities at integer $t$}In the forward limit of $t\rightarrow 0$, the string residue takes a simple, universal form, $R(n,0)=1/n$. Moreover,  $R(n,1) = 1$.
As a consequence, at these values of $t$, $\Omega(z,t)$ itself is just the Riemann zeta function~\footnote{In other words, we have constructed two interesting integral forms of the Riemann zeta function, expressed via the definition of $\Omega(z,t)$,\vspace{-1.5mm}
\begin{equation*}
\begin{aligned}
\zeta(z)&=-\lim_{t\rightarrow0}\frac{1}{2\pi i}\protect\oint_{{\cal C}}\frac{{\rm d}s}{s^{z-1}}\frac{\Gamma(-s)\Gamma(-t)}{\Gamma(1-s-t)}
\\&=-\lim_{t\rightarrow1}\frac{1}{2\pi i}\protect\oint_{{\cal C}}\frac{{\rm d}s}{s^{z}}\frac{\Gamma(-s)\Gamma(-t)}{\Gamma(1-s-t)}. 
\end{aligned}\vspace{-0.5mm}
\end{equation*}
These two forms of $\zeta(z)$ are manifestly the same using the behavior of gamma functions under shifts.},
\be
\Omega(z,0) = \zeta(z+1)\;\;\;{\rm and}\;\;\; \Omega(z,1) = \zeta(z). \label{eq:Omegazeta}
\ee
See the appendix for more relations between $\Omega(z,t)$ and other functions in the literature.

We can write the string residues as polynomials in $t$ using the unsigned Stirling numbers of the first kind,
\be
R(n,t)=\frac{1}{n!}\sum_{k=-1}^{n-1}S_{1}(n,k+1)t^{k}.
\ee
Notably, the residues of the string obey the duality 
\be 
R(n,t) = R(t,n),\label{eq:Rduality}
\ee 
manifest in Eq.~\eqref{eq:R} and motivated on more general dual resonance bootstrap grounds in Ref.~\cite{Cheung:2023adk}. It follows that for nonnegative integer $t=k$, we can write $R(n,k)$ as a polynomial in $n$, which gives us an expression for $\Omega(z,k)$,
\be
\Omega(z,k)=\sum_{n=1}^{\infty}\frac{1}{k!}\sum_{j=-1}^{k-1}S_{1}(k,j+1)n^{j-z}.
\ee
Analytically continuing to all $z$, we have a remarkable identity for $\Omega(z,k)$ as a sum over shifted $\zeta$ functions,
\be
\Omega(z,k)=\frac{1}{k!}\sum_{j=0}^{k}S_{1}(k,j)\,\zeta(z{+}1{-}j).\label{eq:fullposiden} 
\ee
For example, $\Omega(z,4)  =(6\zeta(z)+11\zeta(z{-}1)+6\zeta(z{-}2)+\zeta(z{-}3))/24$.
From the pole at $\zeta(1)$, we see that $\Omega(z,k)$ has poles in the $z$ plane at $z=1,\ldots,k$. 

Meanwhile, for $t$ a negative integer, the level truncation property discussed in Ref.~\cite{Cheung:2024uhn}---that $R(n,t)=0$ when we set $t=-k$ for integer $k < n$---implies that the dual resonant form of the string amplitude can be written as a finite sum,
\be
A(s,-k) = \sum_{n=0}^{k} \frac{R(n,-k)}{n-s}.
\ee
Similarly, level truncation gives $\Omega(z,-k)$ as a finite sum,
\be 
\Omega(z,-k)=\sum_{n=1}^{k}\frac{(-1)^{n+1}(k-1)!}{n!(k-n)!n^{z}},\label{eq:leveltruncation}
\ee
so $\Omega(z,-1)=1$, $\Omega(z,-2)= 1 - 2^{-1-z}$, $\Omega(z,-3)=1-2^{-z} + 3^{-1-z}$, and so on.

\medskip

\mysubsec{Shift relations}Given knowledge of $\Omega(z,t)$ in the entire $z$ plane at some fixed  $t$ with $0 \leq {\rm Re}(t) < 1$, we can uniquely determine $\Omega(z,t)$ for all $z$ at {\it any} fixed $t$ with ${\rm Re}(t) \geq 1$ via a beautiful shift relation. 
We derive the relation as follows. 
By the residue duality in Eq.~\eqref{eq:Rduality} along with level truncation, from which $(n\,{+}\,1)R(n\,{+}\,1,t)=(n+t)R(n,t)$ becomes $(t+1)R(n,t+1)=(n+t)R(n,t)$, Eq.~\eqref{eq:sum} implies a powerful shift relation, 
\be
(t+1)\Omega(z,t+1)  = \Omega(z-1,t) + t\,\Omega(z,t).\label{eq:shift}
\ee
Of course, we can infer this result from the properties of the gamma function, which imply a shift relation for $A(s,t)$ parallel to Eq.~\eqref{eq:shift},
\be 
(t+1)A(s,t+1) = (s+t)A(s,t).\label{eq:Ashift}
\ee
Inputting this observation into the contour integral in Eq.~\eqref{eq:contour}, we arrive at Eq.~\eqref{eq:shift} directly.

Since Eq.~\eqref{eq:shift} tells us $\Omega(z,t+1)$ given the data of $\Omega(z,t)$ for all $z\in\mathbb{C}$, it is then natural to ask the converse: given $\Omega(z,t)$ at arbitrary $z$, can we determine $\Omega(z,t-1)$?
Again using properties of the gamma function, we have $n\,R(n,t-1)=t\,R(n-1,t)$, so from the Dirichlet series form of $\Omega(z,t)$ in Eq.~\eqref{eq:sum}, we have 
\be 
\hspace{-2mm}\begin{aligned}\Omega(z,t-1) & =t\sum_{n=1}^{\infty}\frac{R(n-1,t)}{n^{z+1}} =t\sum_{n=0}^{\infty}\frac{R(n,t)}{(n+1)^{z+1}}\\
 & =1\,{+}\,t\sum_{n=1}^{\infty}\sum_{k=0}^{\infty}\frac{R(n,t)}{n^{z{+}k{+}1}}\frac{(-1)^{k}\Gamma(z{+}k{+}1)}{k!\Gamma(z\,{+}\,1)},
\end{aligned}\hspace{-2mm}
\ee
so we arrive at the negative shift identity,
\be
\hspace{-2mm}\Omega(z,t\,{-}\,1)\,{=}\,1\,{+}\,t\sum_{k=0}^{\infty}\frac{(-1)^{k}\Gamma(z{+}k{+}1)}{k!\Gamma(z\,{+}\,1)}\Omega(z{+}k{+}1,t),\hspace{-2mm}\label{eq:negshift}
\ee
which holds when the sum converges. Comparing with the level truncation identity for negative integer $t$ in Eq.~\eqref{eq:leveltruncation}, we find precise agreement; we also observe numerical agreement for positive integer $t$, comparing with Eq.~\eqref{eq:fullposiden}, in the real $z>0$ regime of convergence.

Together, Eqs.~\eqref{eq:shift} and \eqref{eq:negshift} imply that  $\Omega(z,t)$, viewed as a complex function in the $z$ plane at some arbitrary fixed $t$, is fully specified by knowledge of $\Omega(z,t)$ in the $z$ plane for $t$ in the strip $0\leq {\rm Re}(t)<1$.

\medskip

\mysec{EFT Expansion}We can use $\Omega$ to express the EFT expansion of the string amplitude. As we have noted, from the contour integral definition in  Eq.~\eqref{eq:contour}, when $z=k$ the branch cut along the negative real $s$ axis  goes away, so that $\Omega(k,t)$ is just given by the residue at infinity of $A(s,t)/s^k$.
That is, $\Omega(k,t)$ defines a subtracted dispersion relation, which computes the coefficient of $s^{k-1}$ in the low-energy expansion of the amplitude, so that 
\be
A(s,t) = \sum_{k=0}^\infty  \Omega(k,t)s^{k-1}. \label{eq:AEFT}
\ee
Such EFT expansions, evaluated at $t\geq 0$, are known to satisfy various positivity bounds as a consequence of unitarity for physically consistent amplitudes~\cite{Adams:2006sv,Cheung:2025nhw,Nicolis:2009qm}.

From the definition of the generalized hypergeometric function, we can write an identity for $\Omega(k,t)$,
\be 
\begin{aligned}
{}_{k+2}F_{k+1}\left[\begin{array}{c}
1,\ldots,1,t+1\\
2,\ldots,2
\end{array};1\right] &= \sum_{n=0}^\infty \frac{(t+1)_n}{(n+1)^{k+1}}\frac{1}{n!} \\
=\sum_{n=1}^{\infty}\frac{\Gamma(t+n)}{n^k n!\Gamma(1+t)} &= \Omega(k,t),
\end{aligned}\label{eq:hypergeo}
\ee 
comparing with the Dirichlet sum in Eq.~\eqref{eq:sum}.
Despite this hypergeometric representation, we would like to express $\Omega(k,t)$, which defines the EFT expansion in Eq.~\eqref{eq:AEFT}, in terms of more explicit functions, and without an infinite sum.
In the forward limit, we already know that $\Omega(k,0)=\zeta(k+1)$. 
We will discover that we can actually express $\Omega(k,t)$ in closed form for all $t$.

Let us first make some definitions. 
For an integer $N$, we write a partition $\lambda$ as a list $\lambda_j$ with $\sum_{j=1}^{N}j\lambda_{j}=N$. 
We define a multinomial partition coefficient appearing in Newton's identities, in the notation of Ref.~\cite{AbramowitzStegun}, $M_{2}(k,\lambda)=k!\prod_{n=1}^{k}1/(n^{\lambda_{n}}\lambda_n!)$.
Let us also define a function $h_k(t) = H_{-t}^{(k+1)}$ for all nonnegative integers $k$ as the analytic continuation of the generalized harmonic numbers $H_n^{(k)} = \sum_{j=1}^n 1/j^k$,
\be 
\begin{aligned}
h_{0}(t) &=\psi(1-t)+\gamma \\
h_{k}(t) &=\frac{(-1)^{k}}{k!}\psi^{(k)}(1-t)+\zeta(k+1),\;\;\;k>0,
\end{aligned}
\ee
in terms of the polygamma functions $\psi^{(k)}$ and the Euler-Mascheroni constant $\gamma$.
In terms of $h_k(t)$ and $M_2(k,\lambda)$, we find that we can write $\Omega(k,t)$ compactly as a sum over all partitions of $k$: 
\be 
\Omega(k,t)=-\frac{1}{k!t}\sum_{\lambda}M_{2}(k,\lambda)\prod_{j=1}^{k}(h_{j-1}(t))^{\lambda_{j}}.\label{eq:Omegakt}
\ee
For example,
\be 
\begin{aligned}
\Omega(3,t)  &=-\frac{1}{6t}\bigg[\left(\gamma\,{+}\,\psi(1{-}t)\right)^{3}\,{+}\,\psi^{(2)}(1{-}t)\,{+}\,2\zeta(3) \\& \qquad\;\;\, \,{+}\,\left(\gamma\,{+}\,\psi(1{-}t)\right)\left(\frac{\pi^{2}}{2}\,{-}\,3\psi^{(1)}(1{-}t)\right)\!\bigg].
\end{aligned}\hspace{-1.5mm}
\ee
We see that $\Omega(k,t)$ has poles at all integer values $t=k'\geq k$, as predicted by Eq.~\eqref{eq:fullposiden}. As $t\rightarrow-\infty$, we have $\Omega(k,t)\rightarrow-(\log(-t))^{k}/k!t$.
One can explicitly check that Eq.~\eqref{eq:Omegakt} obeys the shift relation in Eq.~\eqref{eq:shift} and, plugging in positive integer $t$, that it obeys Eq.~\eqref{eq:fullposiden}.

Let us now introduce the multiple zeta functions, which are defined analogously to $\zeta(z)$ in series form,
\be
\zeta(z_{1},\ldots,z_{N})=\sum_{n_{1}>n_{2}>\cdots >n_{N}>0}\frac{1}{n_{1}^{z_{1}}n_{2}^{z_{2}}\cdots n_{N}^{z_{N}}},
\ee
and analytically continued elsewhere~\cite{Akiyama,Matsumoto,Zhao2000}. 
Following the usual notation, we write $\zeta(z,1,\ldots,1)$, where there are $N$ instances of $1$ in a row, as $\zeta(z,\{1\}^{N})$. The standard zeta function in this notation is $\zeta(z)=\zeta(z,\{1\}^{0})$. The function $\zeta(z,\{1\}^{N-1})$ is known as a height 1 zeta function of depth $N$~\cite{Young2023,Young2024}.
The simultaneous expansion in both $s$ and $t$ of the string amplitude has been found to be given in terms of multiple zeta values~\cite{ZagierZerbini,Schlotterer:2012ny,Green:2019tpt},
\be
A(s,t)=-\frac{1}{st}+\sum_{n=0}^{\infty}\sum_{k=0}^{\infty}\zeta(k+2,\{1\}^{n})s^{k}t^{n} .\label{eq:AEFTexp}
\ee
Since the amplitude obeys crossing symmetry, $A(s,t)=A(t,s)$, we have the remarkable duality~\cite{OhnoZagier,Zudilin,KanekoSakata},
\be
\zeta(k+1,\{1\}^{n-1})=\zeta(n+1,\{1\}^{k-1}). 
\ee
In contrast to these results, the resummed form of the EFT expansion in $s$ with $t$ finite, given in Eq.~\eqref{eq:Omegakt} above, was previously unknown.
 
Of course, if we compute the expansion of $\Omega(k,t)$ order by order in $t$, we will reproduce Eq.~\eqref{eq:AEFTexp}.
However, we can obtain a stronger result by expanding $\Omega(z,t)$ in $t$ at complex $z$.
Doing so, we will obtain the multiple zeta function itself evaluated at arbitrary $z$, rather than on the integers.
For example, to compute four $t$ derivatives of $\Omega(z,t)$, we extract the $t^4$ coefficient in $R(n,t)$, which after much careful rearrangement of the sums becomes
\be 
\begin{aligned}
&\partial_{t}^{4}\Omega(z,t)|_{t=0}  
\\&=\sum_{n=1}^{\infty}\frac{1}{n^{z+1}}\bigg[(H_{n-1}^{(1)})^{4}-6H_{n-1}^{(2)}(H_{n-1}^{(1)})^{2}
\\& \qquad\qquad -6H_{n-1}^{(4)}+3(H_{n-1}^{(2)})^{2}+8H_{n-1}^{(3)}H_{n-1}^{(1)}\bigg]\\
 & =\sum_{n_{1}=1}^{\infty}\frac{1}{n_{1}^{z+1}}\Biggl(\sum_{n_2=1}^{n_1-1}\sum_{n_3=1}^{n_1-1}\sum_{n_4=1}^{n_1-1}\sum_{n_5=1}^{n_1-1}\frac{1}{n_{2}n_{3}n_{4}n_{5}} \\&\qquad\qquad\qquad -\sum_{n_2=1}^{n_1-1}\sum_{n_3=1}^{n_1-1}\sum_{n_4=1}^{n_1-1}\frac{6}{n_{2}^{2}n_{3}n_{4}} -\sum_{n_{2}=1}^{n_{1}-1}\frac{6}{n_{2}^{4}}\\&\qquad\qquad\qquad +\sum_{n_2=1}^{n_1-1}\sum_{n_3=1}^{n_1-1}\frac{3}{n_{2}^{2}n_{3}^{2}}+\sum_{n_2=1}^{n_1-1}\sum_{n_3=1}^{n_1-1}\frac{8}{n_{2}^{3}n_{3}}\Biggr)\\
 & =24\sum_{n_{1}=1}^{\infty}\frac{1}{n_{1}^{z+1}}\sum_{n_{2}=1}^{n_{1}-1}\sum_{n_{3}=1}^{n_{2}-1}\sum_{n_{4}=1}^{n_{3}-1}\sum_{n_{5}=1}^{n_{4}-1}\frac{1}{n_{2}n_{3}n_{4}n_{5}}\\
 & =24\,\zeta(z+1,\{1\}^{4}).
\end{aligned}\hspace{-3mm}
\ee
The analogous calculation can be done for any $n$, and one finds 
$\partial_{t}^{n}\Omega(z,t)|_{t=0}=n!\,\zeta(z+1,\{1\}^{n})$.
We therefore have the $t$ expansion of $\Omega(z,t)$, 
\be 
\Omega(z,t)=\sum_{n=0}^{\infty}\zeta(z+1,\{1\}^{n})t^{n}.\label{eq:expand}
\ee
This expression allows us to analytically continue $\Omega(z,t)$ to arbitrary $z$ at fixed $t$ for which the $t$ expansion converges, using the analytic continuation of the multiple zeta functions. 
We can thus indeed view $\Omega(z,t)$ as a $t$-deformation of the Riemann zeta function.

\medskip

\begin{figure*}[t]
\includegraphics[width=\columnwidth]{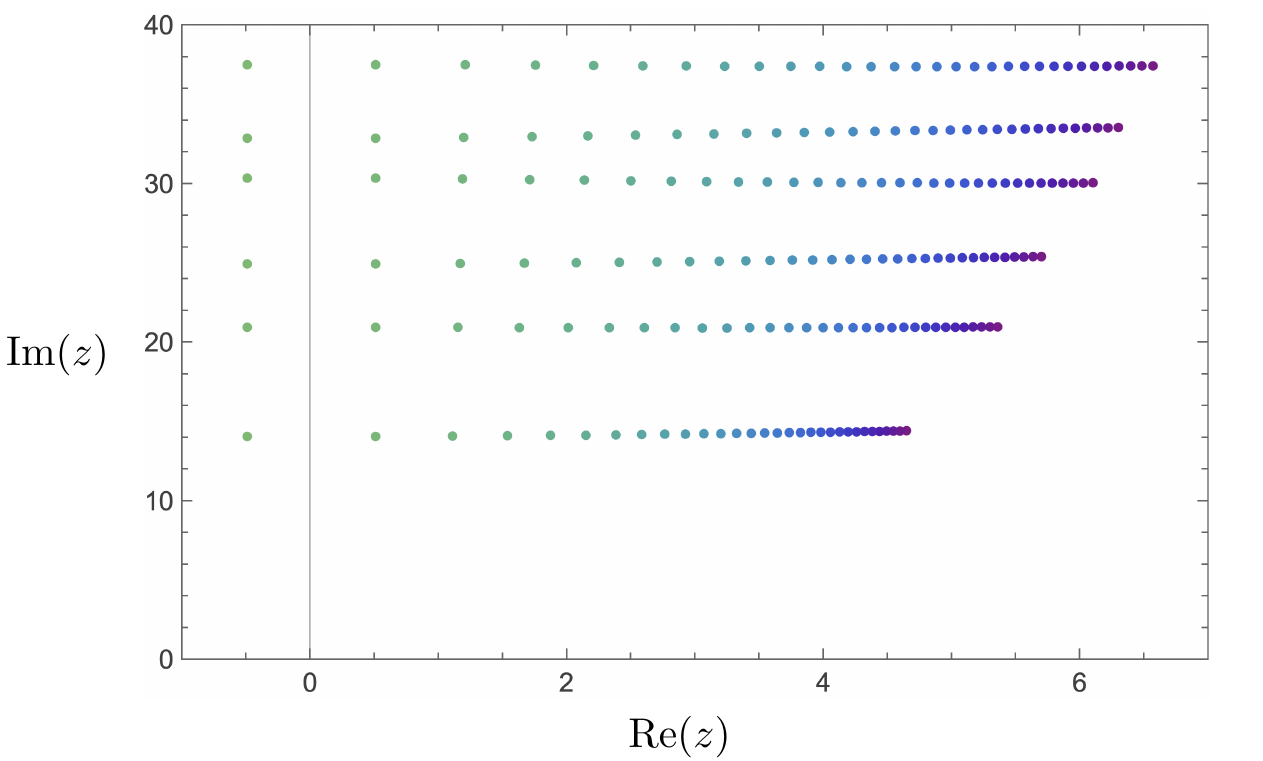}\hspace{3mm}
\includegraphics[width=\columnwidth]{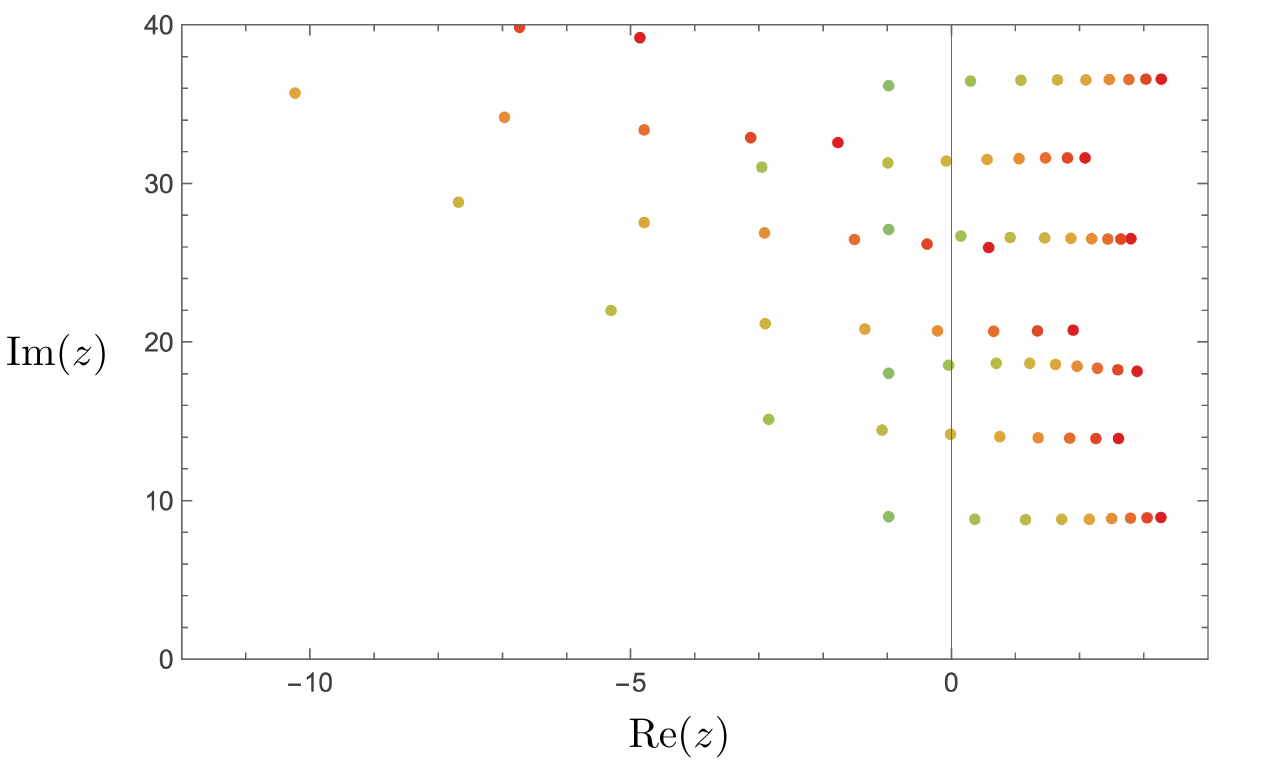}
\caption{Zeros of $\Omega(z,t)$ in the complex $z$ plane for integer $t$. Only zeros off of the real axis are plotted. Formulas are given in Eqs.~\eqref{eq:fullposiden} and \eqref{eq:leveltruncation} for the positive and negative cases, respectively. Left: Positive case, where $t=k$ for $k=0$ (green) through $k=30$ (purple). For the cases $k=0$ and $1$, $\Omega(z,k)$ corresponds to the Riemann zeta function as in Eq.~\eqref{eq:Omegazeta}. Right: Negative case, where $t=k$ for $k=-10$ (red) through $k=-2$ (green). Note that $\Omega(z,-1)=1$, so there are no zeros for $k=-1$.}
\label{fig:zeros}
\end{figure*}

\mysec{Flow of Zeros}Since $\Omega(z,t)$ is a $t$-deformation of the Riemann zeta function, it is interesting to consider the structure of its zeros.
We saw in Eq.~\eqref{eq:integerzeros} that ultrasoftness of the string amplitude gave us the string sum rules in Eq.~\eqref{eq:sumrules}, which imply zeros of $\Omega(z,t)$ at certain integer $z$.
Let us now instead consider the {\it flow}, under $t$-deformation, of the nontrivial zeros of the zeta function. That is, we define some parameterized contour $z(t)$ in the $z$ plane such that 
\be
\Omega(z(t),t)=0 \label{eq:ztzero}
\ee
for all $t$. By Eq.~\eqref{eq:Omegazeta}, $z(0)$ corresponds to a (shifted) zero of the Riemann zeta function, $\zeta(z(0)\,{+}\,1)=0$.
Taking ${\rm d}/{\rm d}t$ of Eq.~\eqref{eq:ztzero}, we can write the flow of the zero,
\be
z'(t)=-\left.\frac{\partial_{t}\Omega(z,t)}{\partial_{z}\Omega(z,t)}\right|_{z=z(t)}. 
\ee
The initial flow is $z'(0)=-\zeta(z(0)\,{+}\,1,1)/\zeta'(z(0)\,{+}\,1)$. Since $\zeta(z(0)\,{+}\,1,1)$ is finite~\cite{Akiyama,Matsumoto,Zhao2000}, $z'(0)$ is finite if and only if the simple zero conjecture for the Riemann zeta function~\cite{Simple} is true.
Since $z'(t)$ is well defined and single valued, the trajectories of two zeros $z_1(t)$ and $z_2(t)$ cannot cross in the $z$ plane.
Similarly, they can merge only if $z_1'(t)=z_2'(t)$, and even then cannot merge permanently, but only instantaneously, provided each $z(t)$ is analytic in $t$.
Since $\Omega(z,-1)=1$ has no zeros, while $\Omega(z,0)=\zeta(z+1)$ and $\Omega(z,-2)=1-2^{-1-z}$ both have infinitely many zeros, it must be that the zeros run away to infinity as we take $t$ through $-1$.
We plot the zeros of $\Omega(z,t)$ for various values of $t$ in Fig.~\ref{fig:zeros}.
We leave the full characterization of the structure of the zeros of $\Omega(z,t)$ in the complex $z$ plane as a function of $t$ to future work.

\medskip

\mysec{Inverse Transform}We saw in Eq.~\eqref{eq:AEFT} that $\Omega(k,t)$ gives the coefficient of $s^{k-1}$ in the EFT expansion of $A(s,t)$. 
We can use this fact to write the string amplitude itself as an $s$-weighted contour over $\Omega(z,t)$ via an inverse Mellin transform, inverting the way in which $\Omega(z,t)$ was defined above as a $z$-weighted contour integral over $A(s,t)$ in the $s$ plane. Concretely, 
\be 
A(s,t)=-\frac{1}{2i}\oint_{\cal C}{\rm d}z\,(-s)^{z-2}\frac{\Omega(z-1,t)}{\sin\pi z},\label{eq:dualcontour}
\ee
where ${\cal C}$ is the same contour, but now in the $z$ plane, as that depicted in Fig.~\ref{fig:contour}, which we used in the definition of $\Omega$ in Eq.~\eqref{eq:contour}.
We can verify Eq.~\eqref{eq:dualcontour} as follows.
Since $\zeta(z,\{1\}^{n})$ is regular for $z>1$, we see that for $t$ such that the series expansion for $\Omega(z-1,t)$ in Eq.~\eqref{eq:expand} converges, the right-hand side of Eq.~\eqref{eq:dualcontour} can be computed by deforming the contour to sum over the poles at positive integer $z$, becoming simply $\sum_{j=1}^{\infty}s^{j-2}\Omega(j-1,t)$.
We recognize from the low-energy expansion in Eq.~\eqref{eq:AEFT} that this sum is precisely $A(s,t)$ itself, thus validating Eq.~\eqref{eq:dualcontour}~\footnote{Note that throughout this section we have taken $(s,t)$ to be in a region where the sums or integrals converge, and then tacitly analytically continued to all $(s,t)$.}. As a consistency check, inputting Eq.~\eqref{eq:shift} into the right-hand side of Eq.~\eqref{eq:dualcontour} and shifting $z$ by $1$---and using that $\Omega(0,t)=-1/t$ for the extra pole at $z=0$---we recover the shift relation for $A(s,t)$ given in Eq.~\eqref{eq:Ashift}, which was implied by the gamma function structure of superstring scattering.

\medskip

\mysec{Discussion}In this paper, we have computed and studied the Mellin transform $\Omega(z,t)$ of the superstring amplitude at fixed Mandelstam $t$, representing a parameterized deformation of the Riemann zeta function.
We found that $\Omega$ exhibits many noteworthy mathematical properties in precise correspondence with physical characteristics of the string amplitude.
Notably, $\Omega(k,t)$ gives a fully $t$-resummed EFT expansion of the string amplitude in powers of $s$.

These results leave many avenues for future study.
Most obvious but challenging is the question of whether the $t$-deformed trajectories of the zeros $z(t)$, which reduce to those of the Riemann zeta function at $t=0$, can shed light on the Riemann hypothesis.
Other promising future directions include generalizing to higher-point, to deformations of the string amplitude~\cite{Cheung:2023adk,Cheung:2024uhn,coon_1969,Maldacena_2022,Haring,Gross:1969db,Cheung:2023uwn,Arkani-Hamed:2023jwn,Cheung:2022mkw}, to the closed string, or to other dispersion relations, e.g., the recent logarithmic derivative dispersion relations of Ref.~\cite{Calisto:2026pvv}, which allow for hard scattering.
Also of interest would be the interpretation of $z$ in the context of Regge theory~\cite{Collins:1977jy}, where the Mellin transform in Eq.~\eqref{eq:contour} replaces the Legendre function with its leading term in a Froissart-Gribov projection of the amplitude in complex angular momentum~\footnote{Similarly, the inverse transform in Eq.~\eqref{eq:dualcontour} is analogous to a Sommerfeld-Watson transform, again keeping only the leading term in the Legendre polynomial.}.
We leave further exploration of these connections to future work.

\medskip\smallskip

\noindent {\it Acknowledgments:} 
I thank Francesco Calisto, Cliff Cheung, Monica Pate, Francesco Sciotti, and Michele Tarquini for comments.
This work was supported by the James Arthur Postdoctoral Fellowship at New York University.

\bibliographystyle{utphys-modified}
\bibliography{string_zeta}

\medskip

\appendix

\mysec{Appendix}While the construction in this paper is novel, involving the fixed-$t$ Mellin transform of the string amplitude---i.e., of the Euler beta integral---and its characterization as a deformation of the Riemann zeta function, for the sake of completeness it is worthwhile to make connections between $\Omega(z,t)$ and other functions in the mathematical literature.

In Ref.~\cite{Young2024}, the multiple Hurwitz-Barnes zeta function of order $r$ was defined as 
\be 
\zeta_{r}(z,a)=\sum_{k_{1}=0}^{\infty}\cdots\sum_{k_{r}=0}^{\infty}(a+k_{1}+\cdots+k_{r})^{-z},
\ee 
and it was shown in Ref.~\cite{Young2014} that 
\be 
\zeta_{r}(z,a)=\sum_{m=0}^{\infty}\binom{m+r-1}{m}(m+a)^{-z},
\ee
where the order $r$ can be analytically continued to the complex numbers. We see that this sum is related to our expression in Eq.~\eqref{eq:sum}; in particular, in Ref.~\cite{Young2024} it was also shown that, setting $a\,{=}\,1$, one has $\zeta_{r}(z)=\sum_{j=0}^{\infty}\zeta(z,\{1\}^{j})(r-1)^{j}$ for ${\rm Re}(z)>1$ and $|r-1|<|z-1|$. Comparing with our series expansion for $\Omega(z,t)$ in Eq.~\eqref{eq:expand} above, we have
\be 
\Omega(z,t)=\zeta_{t+1}(z+1),
\ee
where the order of the Hurwitz-Barnes multiple zeta function has been analytically continued in $t$ for ${\rm Re}(z)>0$ and $|t|<|z|$. 
That is, we have found that the multiple Hurwitz zeta function---which represents a $t$-deformation of the Riemann zeta function---is computed by the dispersion integral of the string amplitude in Eq.~\eqref{eq:contour}. 
As a result, Eq.~(2.11) of Ref.~\cite{Young2024} gives us the relation $\Omega(z,1)=\zeta(z)$, which we found above, and more generally Corollary 1 of Ref.~\cite{Young2024} is equivalent to Eq.~\eqref{eq:fullposiden} above, which yields an expression for $\Omega(z,t)$ as a Stirling-weighted sum of shifted Riemann zeta functions when $t$ is a positive integer.

Further, in Ref.~\cite{Navas2021}, the following binomial-weighted Dirichlet sum  was considered, 
\be 
\mathfrak{B}_{\alpha}^{*}(s)=\sum_{n=0}^{\infty}(-1)^{n} \binom{\alpha}{n}\frac{1}{(n+1)^{s}}.\label{eq:BB}
\ee 
By the reflection formula for the gamma function, we have $n!\,R(n,t) = (-1)^{n-1}\Gamma(-t)/\Gamma(1-n-t)$ from the definition of the residue in Eq.~\eqref{eq:R}, so
\be 
\mathfrak{B}_{-t-1}^{*}(z+1)=\Omega(z,t).
\ee
However, it was not realized in Ref.~\cite{Navas2021} that this function was related to the contour integral in Eq.~\eqref{eq:contour}, nor were further properties explored, such as the connection to the Riemann zeta function. In terms of $\mathfrak{B}^*_\alpha(s)$, the hypergeometric identity in Eq.~\eqref{eq:hypergeo} was found in Ref.~\cite{Navas2021}.

Another related concept is that of the Roman harmonic numbers, given via Knuth's formula as~\cite{Roman} 
\be 
c_{n}^{(k)}=\sum_{j=1}^{n}(-1)^{j-1}\binom{n}{j}\frac{1}{j^{k}}.
\ee
Rearranging the sum and comparing with Eq.~\eqref{eq:BB}, one can see that 
$c_{n}^{(k)}=(n\,{+}\,1)\mathfrak{B}_{n}^{*}(k\,{+}\,1)-\mathfrak{B}_{n}^{*}(k)=(n\,{+}\,1)\Omega(k,-n\,{-}\,1)-\Omega(k\,{-}\,1,-n\,{-}\,1)$.
Using Eq.~\eqref{eq:shift}, we can simplify this relation to 
\be 
\Omega(k,-n)=\frac{c_{n}^{(k)}}{n}.
\ee

In Ref.~\cite{Hasse1930}, the following globally convergent series for the zeta function is given:
\be \zeta(s)=\frac{1}{s-1}\sum_{n=0}^{\infty}\frac{1}{n+1}\sum_{k=0}^{n}\binom{n}{k}\frac{(-1)^{k}}{(k+1)^{s-1}},\label{eq:Hasse}
\ee 
which in terms of our notation above yields the following expression for $\zeta(z+1)$, 
\be 
\Omega(z,0)=\frac{1}{z}\sum_{n=0}^{\infty}\frac{\mathfrak{B}_{n}^{*}(z)}{n+1}=\frac{1}{z}\sum_{n=0}^{\infty}\frac{\Omega(z-1,-n-1)}{n+1}.\label{eq:Hasse2}
\ee 
Using the expression for $\Omega(z-1,-n-1)$ from Eq.~\eqref{eq:leveltruncation}, we find that the sum in the above expression is given by
\be \hspace{-0.5mm}
\sum_{n=0}^{\infty} \! \frac{\Omega(z-1,{-}n{-}1)}{n+1} \! =\! \sum_{n=0}^{\infty}\sum_{k=1}^{n+1}\! \frac{(-1)^{k{+}1}}{(n{+}1)^2 k^{z{-}1}} \binom{n{+}1}{k}.\hspace{-1mm}
\ee 
We recognize from Eq.~\eqref{eq:Hasse} that this final expression equals $z\,\zeta(z+1)$, as required by Eq.~\eqref{eq:Hasse2}. Thus, our finite sum in Eq.~\eqref{eq:leveltruncation} is consistent with Eq.~\eqref{eq:Hasse}.

\end{document}